# Entropic measure to prevent energy over-minimization in molecular dynamics simulations

J. Rydzewski, R. Jakubowski and W. Nowak[1]

Institute of Physics, Faculty of Physics, Astronomy and Informatics, Nicolaus Copernicus University, Grudziadzka 5, 87-100 Torun, Poland

This work examines the impact of energy over-minimization on an ensemble of biological molecules subjected to the potential energy minimization procedure in vacuum. In the studied structures, long potential energy minimization stage leads to an increase of the main- and side-chain entropies in proteins. We show that such over-minimization may diverge the protein structures from the near-native attraction basin which possesses a minimum of free energy. We propose a measure based on the Pareto front of total entropy for quality assessment of minimized protein conformation. This measure may help in selection of adequate number of energy minimization steps in protein modelling and, thus, in preservation of the near-native protein conformation.

## I. INTRODUCTION

The protein structure refinement is routinely performed using potential energy minimization methods in the field of crystallography, structure prediction and molecular dynamics[1,2]. Protein energy is usually expressed as a functional form of atomic coordinates, called a force field (FF)[3]. In recent years, many energy minimization techniques has been developed[4-6]. The conjugated gradient method (CG) is probably the most popular one [7]. The goal of energy minimization is as follows: a model protein structure is needed, free of steric clashes[8,9], with potential energy preferably lower than that of a crystal structure[10] and lying as close to the native structure as possible in terms of root-mean-square deviation (RMSD). Classical FFs (i.e. CHARMM, AMBER, GROMOS and OPLS), based on potential energy, do not contain entropic terms. Thus, a local minimum of energy found after numerical minimization does not necessarily correspond to the most thermodynamically stable structure. The thermodynamically stable structure resides in a minimum on the free-energy ($\Delta G$) surface[11]. The structure having the lowest free-energy should correspond to a native protein structure[12-19].

Success of molecular dynamics (MD) simulations depends partially on quality of the initial structure. The structure from the energy minimization step is routinely used for further thermalization, equilibration and production runs. Therefore, it is particularly troublesome to observe over-minimization, since its effects may be hidden or decreased due to subsequent preparatory steps. However, since over-minimization can perhaps diverge the protein structures from the near-native attraction basin and, thus, disrupts protein biological functions, it is important to examine this methodological issue more closely. Given the notion that energy minimization techniques based on classical FFs generally do not preserve the native structures of proteins as observed by x-ray crystallography[20], one can ask how to improve this initial step of protein preparation for MD simulations and assume that the best starting point for MD simulations would be a structure that approximate the native state as accurately as possible[20].

---

[1] Email to corresponding author: wiesiek@fizyka.umk.pl

## II. THE SIMULATION PROTOCOL

Energy minimization is considered converged if the norm of the energy gradient is close to zero and further iterations do not lower energy more than a given threshold. In preparatory steps some codes, i.e. NAMD, CHARMM, accept user-provided number of CG iterations as the only criterion for the termination of the structure refinement protocol. In manuals, one can find a caveat that this protocol results often in over-minimization which renders a protein structure unphysical (in contrast to physical "protein-like" conformations[21]). However, an explanation of over-minimization process remains unclear. The protein adopts its near-native structure by the thermodynamically controlled process, involving enthalpic contributions of non-covalent bond formations, entropic and hydrophobic effects. It is an accepted paradigm that compensation of such driving forces leads to the minimum of free-energy ($\Delta G$) calculated for a given protein[22]. Therefore, in order to monitor the effect of the protein potential energy minimization on its near-native structure conservation, calculations should incorporate a $\Delta G$ difference ($\Delta\Delta G$).

### A. Generation of the protein ensemble

With this in mind, we minimized energies of a large set of protein structures (70 template proteins and nearly 800 derived perturbed structures) with the well-established CG method[7]. To analyse thermodynamic character of changes in protein structures occurring on paths toward the local potential energy minima, we monitored free energy changes, $\Delta\Delta G$, as predicted by the FoldX program[23]. Moreover, our calculations were also aimed at quantifying possible free-energy destabilization effects of protein structures during over-minimization and finding a method for determination of the optimal number of CG energy minimization steps before routine MD simulations. In order to address the problem, we generated a large ensemble of the protein decoy structures. We selected 70 protein structures from Protein Data Bank[24] with the x-ray resolution better than 2.5 Å (Supplementary Information, List 1). Of these 70 proteins, 15 were picked because of their known rugged energy landscape (i.e. leu-enkephalin, interleukin-1β, barnase) or relevance in medicine (i.e. neurexin, neuroligin). The remaining 55 proteins were chosen from the pdbfinder database[25] according to the criterion of being wild-type and monomeric. All protein ligands, if present in the chosen crystal structures, were discarded. Each structure was a template for generation of a larger set of input structures. The position of each atom in the template protein was shifted by a random vector in such a way that RMSD between the decoy and input structures did not exceed 1Å. Each template was used to generate additional 10 decoy protein structures. This procedure gave an ensemble consisting of nearly 800 starting protein structures representing both, nearly native and randomly perturbed biomolecules.

### B. Energy minimization

Potential energy of each structure from the ensemble was minimized in vacuum using the NAMD program[26] and CHARMM27 force field[27]. We employed the CG method, since it is one of the most popular minimization algorithm. First, we have performed 10k steps of energy minimization to explore long-term energetic trends. Since the most substantial changes were observed in the first 1k steps, in this article we focus on this shorter minimization phase. Proteins' snapshots were recorded every 10 CG steps. Every protein structure from the minimization process was subjected to calculations of free-energy using the FoldX. The whole ensemble's energy minimization paths were subsequently examined in terms of free-energy decomposition, in order to check whether the postulated over-minimization of the protein structures could be noticed and quantified.

**C. Energy decomposition**

As already mentioned, derivation of the ΔΔG distributions for all steps during energy minimization for each protein structure was done using the experimentally validated FoldX algorithm[23,28]. Although FoldX ΔG values are predictions of limited accuracy, the examination of the ΔΔG distributions for a large set of proteins during energy minimizations provided details on the protein structure refinement process. ΔΔG values were estimated using temperature set to 298 K and pH equal to 7. One should note that our PDB structures were not optimized by FoldX, since the structure refinement process was performed solely in terms of potential energy minimization, with the NAMD simulation package using CG, therefore just a single point ΔG calculations were performed. The empirical force field of FoldX consists of a linear combination of the following energy terms: van der Waals forces, solvation of hydrophobic and polar groups, backbone and side-chain hydrogen bonds, electrostatic interactions, side-chain and main-chain entropies, and torsional and backbone clashes. The details concerning the calculation of these terms are discussed elsewhere in detail[23].

**III. RESULTS AND DISSCUSION**

**A. Energy decomposition**

The results show that the main positive contribution to the total free energy comes from main-chain entropy, $S_M$, side-chain entropy, $S_S$ and solvation energy for polar groups. On the other hand, the negative contribution is due to the following enthalphic terms: hydrophobic solvation, van der Waals and hydrogen bonds energy. From a theoretical point of view, the compensation of the entropic and enthalphic terms drives the protein structure into thermodynamic stabilization[29]. Our results show that during the minimization from the free-energy minimum to the last step free energy increased by more than 20% (FIG. 1). Our estimations indicate that the main contribution to the total free energies of over-minimized protein structures is due to a sudden rise of the side-chain and main-chain entropies. Under normal conditions, an increase of the entropic energy should be compensated with a decrease of the enthalpic energy of non-covalent bonds. Moreover, such long minimization leads to an increase of the van der Waals energy clashes caused by close packing of proteins' residues what stands against energy minimization aims.

To study an impact of the total entropy on the ensemble of structures during minimization, we calculated the total free-energy, ΔG, and the free-energy with excluded entropies, $ΔG_{-S}$ (FIG. 2). Energy decomposition shows that the ensemble of structures displays two stages during energy minimization (FIG. 2). The first stage is characterized by a steep decrease of ΔG, after which the ensemble of structures reaches a low free energy basin. The global ΔG and $ΔG_{-S}$ minima are reached in this stage at approximately 210[th] step. When the total entropy is not taken into account of ΔG, the ensemble overcomes a free-energy barrier and arrives at the

$\Delta G_S$ plateau. A different scenario is displayed if the total entropy is taken into account, although the first stage proceeds in the similar way. After the first stage, over-minimization manifests leading to an increase of $\Delta G$.

Current studies often report protocols with more than 5000 energy minimization steps[30–35]. However, our results indicate that performing such long minimizations may lead to thermodynamic destabilization of proteins which means that these protein structures may diverge from the near-native attraction basin. As a result, using an unphysical protein structure in MD is perhaps not a problem only if a full conformational space is sampled, however, such ergodicity requirement is rarely meet in practice. Thus, such an over-minimized initial structure may lead to unphysical results of simulations and to false conclusions, since non-optimum basins of free energy surface are sampled in limited time MD simulations.

## B. Pareto front of total entropy

There are warnings in the program manuals that over-minimization should be avoided[36]. However, to the best of our knowledge, there is no concise criterion defining clearly when the minimization process should be stopped. We show that it is possible to propose such a concise measure based on the total entropy. The entropic effects are incorporated in the free-energy calculations in two parts: the main-chain entropy and the side-chain entropy. The first one is the entropic cost of fixing the backbone in the folded state. In other words, this term is dependent on the intrinsic tendency of amino acids to adopt certain dihedral angles. The second one is the entropic cost of fixing side chains in a particular conformation, in such a way that the side-chain entropy of a completely buried residue is assumed to be 0.

Zhang *et al.* have shown that side-chain entropy is crucial in favoring the near-native proteins structures[37]. Therefore, it is important to study the preservation of the near-native states during the minimization protocol by incorporating side-chain entropy. Figure 3 depicts how both the side-chain and main-chain entropies are larger for proteins with more residues. It is understandable since longer proteins may have bigger apolar surfaces which indicate larger hydrophobic effects. Protein residues which are apolar and exposed at the protein exterior surface, contribute to the side-chain entropy. On the other hand, the main-chain entropy increase is caused either by the accessibility of the main-chain atoms or the energy contribution of hydrogen bond interactions made by the corresponding residue or its direct neighbors[23].

Moreover, Figure 3 inset illustrates that the minimization process of each protein from the ensemble appears to be a two-fold process when addressing the relation between $S_M$ and $S_S$. Firstly, a fast decrease of $S_M$ with a little change in $S_S$ can be observed. This stage moves the conformation of a particular protein from the crystal structure to the stabilized protein conformation, near the protein structure with the lowest $S_M + S_S$. Secondly, the over-minimization occurs, and manifests itself by alternate increases in $S_M$ and $S_S$, until the minimization process is over. This final growth in total entropy is not compensated by the enthalpic energy terms, which leads to approximately 20% increase of the total free energy of the ensemble. We calculated the Pareto front of total entropy $S_M + S_S$, which consists of all protein structures that are characterized by the minimum of $S_M + S_S$ during the minimization process. As one can observe, the thermodynamically stable structures are closer (in terms of distance in the entropy configuration space) to the Pareto front than the energy over-minimized protein structures (FIG. 3).

We postulate, therefore, a measure to check whether a protein conformation is optimally minimized: (*i*) energy minimization should be run for 1000 CG steps, (*ii*) correlation between the main-chain and side-chain entropies should be examined and protein conformations (acquired during minimization) classified in the following categories: a crystal structure, a minimized structure, a thermodynamically stable structure, a structure characterized by the lowest total entropy $S_M + S_S$ (the Pareto front), and (*iii*) the thermodynamically stable conformation should be chosen as the best starting point for further MD simulations if and only if it is closer to the Pareto front of total entropy than the previous conformation achieved upon energy minimization. In other words, if the last conformation is far away from the Pareto front in the entropy configuration space, it means that this structure is over-minimized and should be omitted in the next structure's refinement stages.

**C. Over-minimization**

We estimated that the most probable minimization step in which our ensemble of the protein structures exhibits the free energy minimum is the $220^{th}$. Nearly 70% of the investigated protein structures reach their native states after performing only 220 steps of energy minimization (FIG. 4). It is clear that over-minimization is particularly unwanted when it leads to protein structures that are not thermodynamically stable. The native states of the proteins, as observed by x-ray crystallography, are not only not preserved, but also driven away from their near-native conformations to "squeezed" structures characterized by high total entropy. We hypothesize that such a lack of entropy compensation is caused by "squeezing" the proteins structures: some hydrophobic residues are pushed out to the protein exterior. Likewise, this process may result in an increase of the van der Waals clashes between atoms. The over-minimization stage leads to an increase of the van der Waals clashes by nearly 20%, which is unexpected during the energy minimization procedure (FIG. 1, 2). One expects that energy minimization increases any too short distances between atoms. Moreover, a formation of additional hydrogen bonds between the side-chains of proteins indicates that the protein conformations are particularly densely-packed during the phase of over-minimization process.

Furthermore, one can observe that there is no linear correlation between the number of minimization steps required to converge to the thermodynamically stable conformation and the number of residues in the protein (FIG. 4, inset). Therefore, it is impossible to optimize the length of the energy minimization protocol using only the size of a given protein – the refinement also strongly depends on the amino acid sequence. This conclusion is in agreement with many previous investigations emphasizing the role of amino acid sequences for the formation of specific secondary structure elements. Thus, the shape of a free-energy landscape is both sequence and length dependent[38].

**IV. CONCLUSIONS**

The following conclusions can be drawn from the foregoing discussions:

a) We show that over-minimization manifests itself mostly in increasing the side- and main-chain entropies, therefore, it is of high importance to incorporate FF with entropic contributions.
b) We propose a quantitative measure based on the Pareto front of total entropy in order to help selection of the CG minimization length and to limit the deviation of the modeled protein structures from the near-native attraction basin.

Taking into consideration the importance of the protein structure refinement in the field of crystallography, structure prediction and molecular dynamics, the foregoing studies of over-minimization should be meaningful not only for detailed understanding of the free-energy landscape in preferring the near-native protein conformations but also for successful development of more advanced force fields.

**ACKNOWLEDGMENTS**

R. Jakubowski acknowledges support from "Krok w przyszlosc – V edycja" scholarship from the Marshal of Kuyavian-pomeranian voivodeship. We would like to thank for the computer time allocated by the Interdisciplinary Centre for Modern Technologies (NCU), and E. Ratajczak and A. Roszak for critically reading the manuscript.

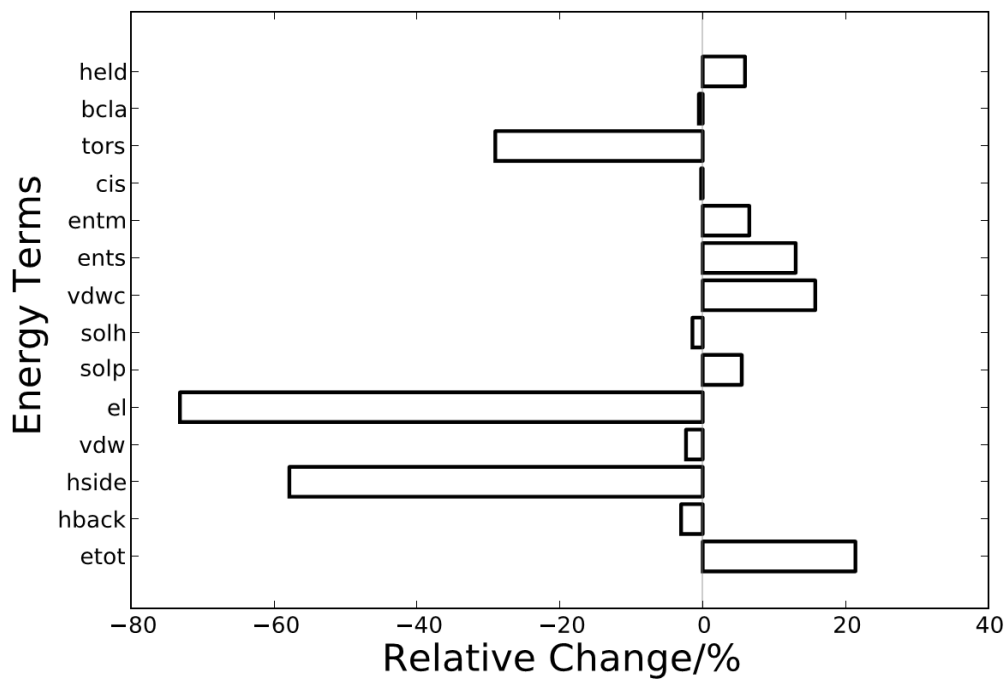

FIG. 1. The relative change $\Delta\Delta G_R$ of free-energy terms of the protein ensemble during minimization from the most thermodynamically stable conformations $\Delta G_S$ to the final conformations after minimization: $\Delta\Delta G_R = (\Delta G_F - \Delta G_S)\Delta G_S^{-1}$. During this part of minimization energy increased by more than 20%. This result shows that performing too many minimization steps – after achieving the free-energy minimum in the $210^{th}$ step – leads to the over-minimized protein structures. Our estimations indicate that the main contribution to the unphysically over-minimized protein structures' total energies is due to a sudden rise of the side-chain and main-chain entropies. Under normal conditions, an increase of the entropic energy should be compensated with a decrease of the enthalpic energy of non-covalent bonds. Moreover, such long minimization leads to an increase of the van der Waals energy clashes caused by close packing of proteins' residues what stands against energy minimization aims. In this figure the following abbreviations for free energy terms are used: etot – total, hback – backbone hydrogen bonds, hside – side-chain hydrogen bonds, vdw – van der Waals, el – electrostatic, solp – solvation of polar residues, solh – solvation of hydrophobic residues, vdwc – van der Waals clashes, ents – side-chain entropy, entm – main-chain entropy, cis – peptide bond in cis conformation, tors – torsional, bcla – backbone clashes, held – helix dipole.

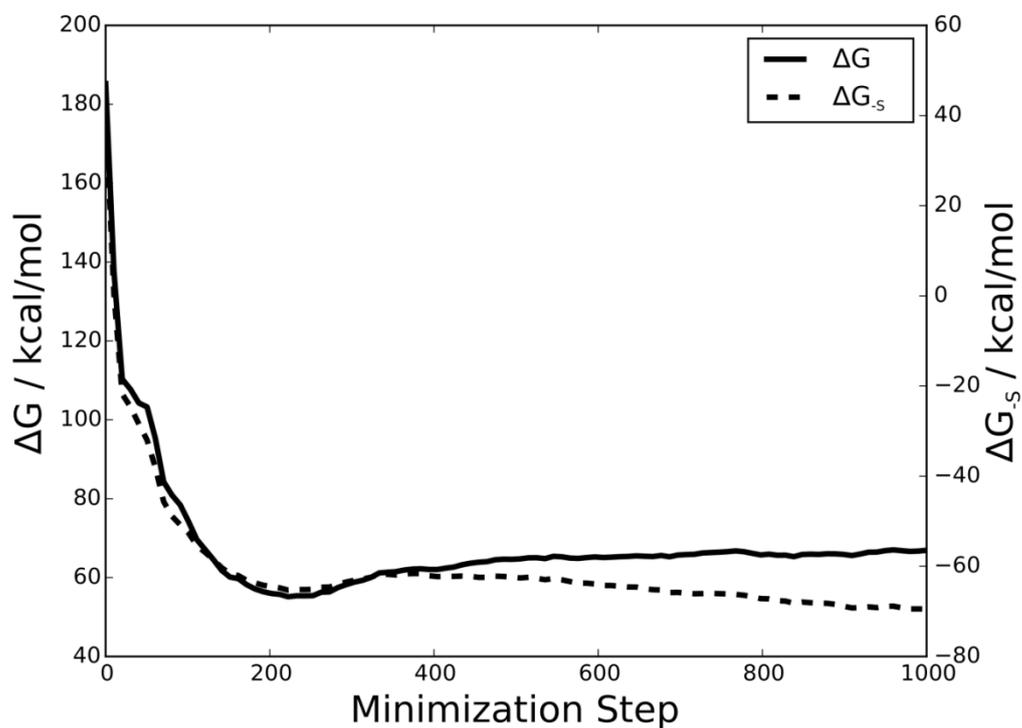

FIG. 2. Impact of the total entropy on the ensemble of structures minimization. The solid line corresponds to the averaged full free energy calculations, whereas dashed line presents the same results with excluded entropic contributions. Up to 300$^{th}$ step similar tendencies can be observed, as in both cases energy rapidly decreases until 210$^{th}$ step and then it slightly grows. After this stage omitting entropic contributions leads to an impression that further minimization brings continuation of ΔG decrease, while in fact an accumulation of entropic effects causes increase of free energy.

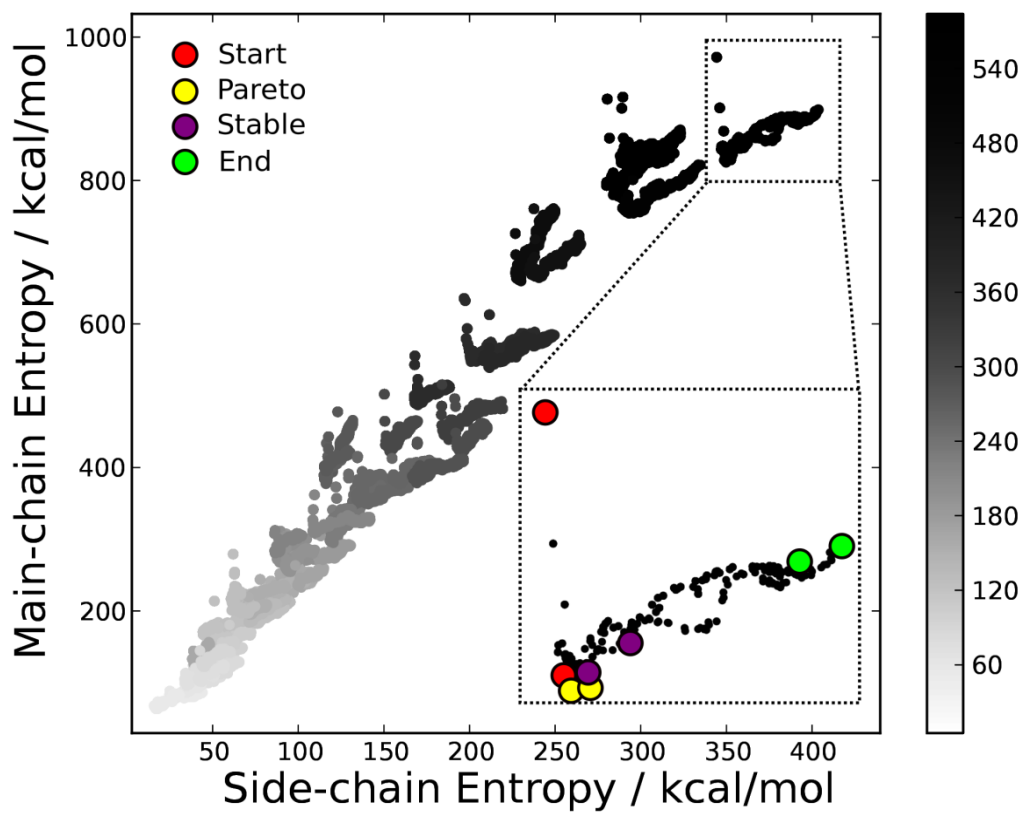

FIG. 3. The side-chain entropy $S_S$ and main-chain entropy $S_M$ relation. The grey scale indicates the number of protein residues. The side-chain entropy is evaluated with the following formula: $S_S = -R \sum_i P_i \ln(P_i) + R \ln(N)$, where the summation occurs over all preferred conformational zones of a particular residue type, $R$ is the gas constant, $P_i$ is the probability of the residue type being in a preferred conformational zone, and $N$ is the additional number of the forbidden rotameric states (with the assumption that all these states have equal probabilities). $S_M$ is derived from statistical analysis of the phi–psi angles distribution of a given amino acid as observed in crystal structures. The inset depicts the positions of the crystal structure of a particular protein, a starting protein structure (red circles), a stabilized protein structure (blue diamonds), an over-minimized protein structure (green triangles) and a protein structure (yellow squares) with the lowest corresponding sum of entropies $S_S$, $S_M$ – Pareto front of total entropy of two protein structures.

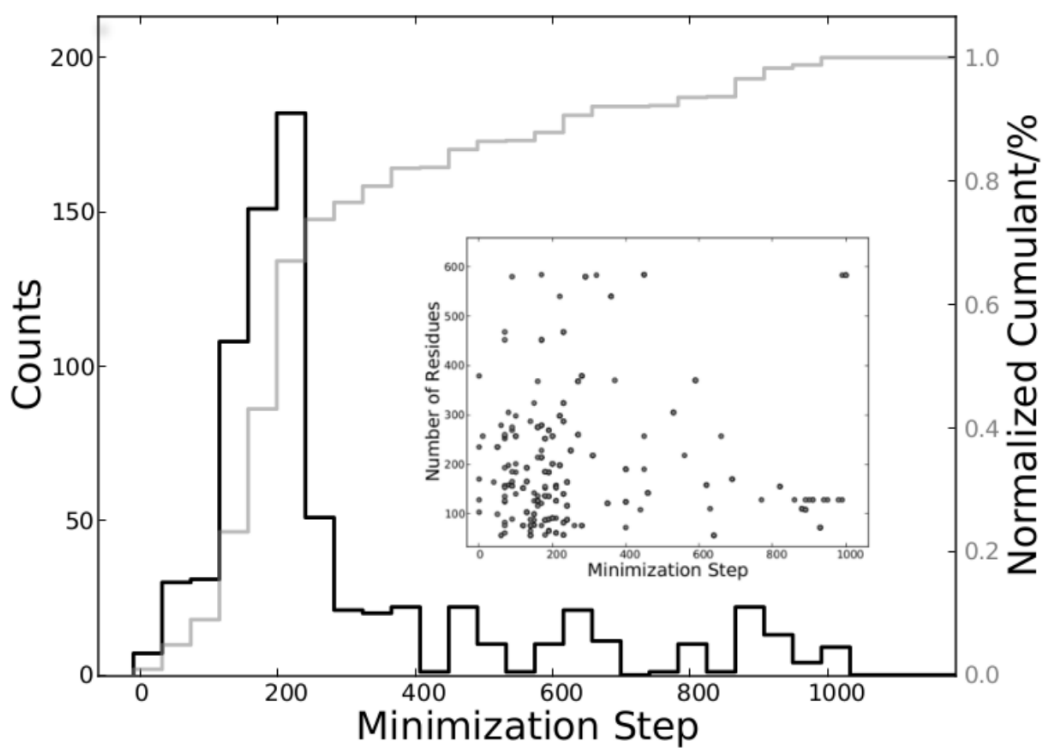

FIG. 4. Histogram (black) of a number of minimization steps leading to the most thermodynamically stable protein conformation for the conjugated gradient method. The normalized cumulant is shown in grey. Considering the 1000-steps minimization protocol, a vast majority of proteins reach their lowest free energy conformations after only 220 steps (about 70%). The occurrences of the most stable conformation during potential energy minimization procedure are counted in 30 bins. We show that over-minimization (in this case, minimization after 220 step) is not necessary and may lead not only to a waste of computational time, but also to unphysically quenched protein conformations. The inset scatter plot shows the lack of correlation between the minimization step number in which a given protein achieves stability and the number of residues in the protein.